\documentclass[reprint,
amsmath,amssymb,
aps,
pre]{revtex4-2}
\usepackage[T1]{fontenc}
\usepackage[latin9]{inputenc}
\usepackage{graphicx}
\usepackage{esint}

\makeatletter
\usepackage{xcolor}
\usepackage{bm}
\usepackage{babel}

\makeatother

\begin{document}
\title{The classical-quantum disproportionation transition and magnetic ordering in RNiO$_3$ nickelates}
\author{A.S. Moskvin}
\affiliation{Ural Federal University, Ekaterinburg, Russia}
\affiliation{M.N. Mikheev Institute of Metal Physics, Ural Branch, Russian Academy of Sciences, Ekaterinburg, Russia}
\author{Yu.D. Panov}
\affiliation{Ural Federal University, Ekaterinburg, Russia}
\begin{abstract}
The insulator-quasi-metal (bad metal) transition observed in Jahn-Teller (JT) magnets orthonickelates RNiO$_3$ (R = rare earth, or yttrium Y) is considered a canonical example of the Mott transition, traditionally described in the framework of Hubbard's $U-t$ model. However, in reality, the insulating phase of nickelates is the result of charge disproportionation (CD) with the formation of a system of spin-triplet ($S = 1$) electron [NiO$_6$]$^{10-}$ and spinless ($S = 0$) hole [NiO$_6$]$^{8-}$ centers, equivalent to a system of effective spin-triplet composite bosons moving in a nonmagnetic lattice. The effective CD-phase Hamiltonian takes into account local ($U$) and nonlocal ($V$) correlations, and the transfer of composite bosons ($t_b$). Within the framework of the effective field approximation, we have shown the existence of two types of CD phases: the high-temperature classical paramagnetic CO-phase of charge ordering of electron and hole centers, and the low-temperature magnetic quantum CDq phase with charge and spin density transfer between electron and hole centers, with ''uncertain valence'' [NiO$_{6}$]$^{(9±\delta)-}$ ($0 \le \delta \le 1$) and spin density $(1 \pm \delta)/2$ NiO$_6$-centers. In the classical CO phase, spin-triplet electron centers are surrounded by the nearest nonmagnetic hole centers, which ''turns off'' the strong superexchange interaction of the nearest neighbors. The magnetic ordering in the quantum CDq phase is determined by a strong traditional superexchange and an unusual bosonic double exchange mechanism.

\end{abstract}
\maketitle

\section{Introduction}

Nickelates RNiO$_3$ (R = rare earth or Y) demonstrate extremely unusual electrical and magnetic properties, this is primarily a poorly pronounced first-order metal-insulator transition (MIT) observed in orthorhombic RNiO$_3$ (R = Lu, $\ldots$, Pr) with cooling below $T_{\mbox{{\footnotesize MIT}}}$ in a range from 130 K for Pr to $\sim$ 550-600 K for heavy rare earths [1-3]. This electrical feature opens wide opportunities for creating temperature-sensitive and resistive switches and other devices, including neuron-spin logic and resistive randomaccess memory [2]. The nature of MIT in nickelates has been a kind of challenge for condensed matter physics during more than three decades. Traditional approach to describing a phase transition with a dramatic change of magnitude and temperature dependence of resistance in nickelates implies that a spontaneous ''metal-insulator'' order-order phase transition is implemented from a hightemperature phase of the coherent Fermi liquid to a lowtemperature charge ordered insulating phase. The Hubbard model [4] is a typical theoretical MIT model and is described by the following Hamiltonian in the simplest case
\begin{equation}
	\hat{H}_{Hub} = - t \sum_{\left\langle i,j \right\rangle \sigma}
	\left( \hat{c}_{i,\sigma}^{+} \hat{c}_{j,\sigma}^{} + h.c. \right) 
	+ \frac{U}{2} \sum_{i} \hat{n}_{i,\sigma} \hat{n}_{i,-\sigma} ,
\end{equation}
that considers the single-particle kinetic energy defined by the effective transfer integral $t$ and local correlations defined by the effective parameter $U$. Transfer integrals are usually within a tight-binding approximation using the DFT (density functional theory) calculation scheme, and $U$ is treated as an adjustable parameter for $d$-electrons. With a quite large $U$ compared with the bandwidth $W \approx 2zt$ ($z$ is the number of the nearest neighbors), i. e. with strong correlations, the ''metal-insulator'' transition is possible. This transition with a certain final value of $U$ was named the bandwidth control (BC)-MIT) [4].

Despite a huge popularity of the Hubbard model and traditional DFT band approaches to describing the ''metal-insulator'' transition that imply any consideration of local correlations for some ''parent'' metallic Fermi liquid phase, they don't provide any adequate description of the real situation in nickelates. In our view, statements of many authors that RNiO$_3$ are ideal candidates for studying BC-MIT are erroneous. Indeed, a high-temperature conducting ''metal-like'' phase in JT magnets differs fundamentally from traditional metals --- extremely low density of carriers, high effective carrier mass and low diffusion capability, low DC conductivity with non-typical temperature dependence, violation of the Mott-Ioffe-Regel criterion, incoherent behavior of charge carriers, observed variation of carrier concentration and sign with temperature variation, clear signs of hopping polaron conductivity, temperature-dependent paramagnetism, detection of electron/hole centers typical of a low-temperature disproportionation phase [4,6-16], which provided a basis for a general term --- ''bad'' or strange metal. Conductivity phenomenon of such bad metals is one of the central problems in the condensed matter physics.

In a low-temperature insulating phase, orthonickelates demonstrate more or less clear signs of charge and bond disproportionation with two types of Ni centers corresponding to large and small NiO$_6$-octahedra, and also a magnetic phase transition to an antiferromagnetic structure that hasn't been known before for perovskite 3$d$-compounds and is defined by the propagation vector $\mathbf{Q}_{AFM} = (1/2, 0, 1/2)$ in orthorhombic coordinates [1,2].

\begin{table*}
\begin{tabular}{p{5em}p{4em}p{6em}p{7em}p{5em}p{6em}}
$d$-center &	Ion	& Cluster&	Pseudospin projection &	Spin &	Orb. state \\ \hline
$d^8$ &	Ni$^{2+}$ &	[NiO$_{6}$]$^{10-}$ &	$M =-1$ &	$S = 1$ &	$t_{2g}^{6}	e_{g}^{2}; \; ^{3}A_{2g}$ \\
$d^7$ &	Ni$^{3+}$ &	[NiO$_{6}$]$^{9-}$ &	$M = 0$ &	$S = 1/2$ &	$t_{2g}^{6}	e_{g}^{1}; \; ^{2}E_{g}$ \\
$d^6$ &	Ni$^{4+}$ &	[NiO$_{6}$]$^{8-}$ &	$M =+1$ &	$S = 0$ &	$t_{2g}^{6}	e_{g}^{}; \; ^{1}A_{1g}$	\\
\end{tabular}	
	\caption{Pseudospin, spin and orbital structure of three charge centers NiO$_6$ in RNiO$_3$}
\end{table*}

Now few people doubt that the low-temperature insulating phase of nickelates results from the charge disproportionation (CD), however, the question of CD phase structure, key interactions and effective Hamiltonian remains open.	This work addresses	a charge ''orthonickelate'' model [17,18] within the charge triplet and pseudospin $\Sigma=1$ formalism model and shows that consideration of local and non-local correlations, and of two-particle transfer allows the nature of MIT, CD phase structure and known phase $T$-R diagram of RNiO$_3$ orthonickelates to be explained.

\section{Charge triplet model}

Following the remarkable idea of Rice and Sneddon [19] developed by us for two-dimensional cuprates and other JT magnets [20-24], we propose a generalized model of effective charge triplets for describing the electronic structure and phase diagrams of RNiO$_3$, that implies addressing a high-symmetry	''parent'' configuration with ideal NiO$_6$ octahedra, low-energy state of which is formed by the charge triplet [NiO$_6$]$^{10-,9-,8-}$ (nominally Ni$^{2+,3+,4+}$) with various spin and orbital ground states. We associate three charge states of the NiO$_6$ cluster with three pseudospin $\Sigma=1$ projections and use the known spin algebra and other methods, that are well-proven for spin magnets, to describe charge degrees of freedom of nickelates in a ''coordinate'' representation instead of a traditional one for models based on the DFT $\mathbf{k}$-representation. Pseudospin, spin and orbital structure of the charge NiO$_6$ centers in nickelates is shown in the Table 1. In the simplest approximation below, we neglect both a probable difference in the $d$-$p$-structure of single-particle $t_{2g}$- and $e_g$-states for different components of the charge triplet, and a contribution of the inactive fully occupied $t_{2g}^6$-shell with $S = 0$ to various spin and orbital interactions.

Formally, the local pseudospin $\Sigma=1$ implies eight (three ''dipole'' and five ''quadrupole'') independent operators and corresponding local charge order parameters. In irreducible components, they are
$$
\Sigma_0=\Sigma_z; \; \Sigma_{\pm}=\mp \frac{1}{\sqrt{2}}(\Sigma_x\pm i\Sigma_y); 
$$
$$
\Sigma_z^2; \; \Sigma_{\pm}^2; \; T_{\pm}=\frac{1}{2}\{\Sigma_z,\Sigma_{\pm}\} \, .
$$
$n_{e_g}$=\,1\,-\,$\langle {\hat \Sigma}_z\rangle$ is the local mean number of $e_g$-electrons,  $\Delta n$\,=\,$\langle {\hat \Sigma}_z\rangle$ defines the deviation from half-occupancy. Operators $P_0=(1-\Sigma_z^2)$; $P_{\pm}=\frac{1}{2}\Sigma_z^2(1\pm\Sigma_z)$  are actually projection operators to charge states with pseudospin projection $M = 0, \pm1$, respectively, and the mean numbers $\langle P_0\rangle$, $\langle P_{\pm}\rangle$, are actually the local densities for corresponding charge states.

The operators $\Sigma_{\pm}$ and $T_{\pm}$ change the pseudospin projection by $\pm$1. The operators $\Sigma_{\pm}^2$ change the pseudospin projection by $\pm$2, therefore they may be treated as creation/ annihilation operators for an effective composite boson. The corresponding local mean numbers $\langle \Sigma_{\pm}\rangle$, $\langle T_{\pm}\rangle$, $\langle \Sigma_{\pm}^2\rangle$ will describe various ''off-diagonal'' charge order alternatives, in particular, coherent metallic and superconducting states.

Taking into account spin and orbital states for charge components, we should expand the local Hilbert space to a ''pseudospin-orbital-spin octet''
$$
|\Sigma M;\Gamma\mu;Sm\rangle = |1M;\Gamma\mu;Sm\rangle \, ,
$$
($\Gamma = A_{1g}, A_{2g}, E_g$ is the irreducible representation of the local point group) including the spin-orbital JT quartet $|10;E_g\mu;\frac{1}{2}\nu\rangle$ with $M = 0$ and spin-charge quartet with $M=\pm1$, including the singlet  $|1+1;A_{1g}0;00\rangle$ and triplet $|1-1;A_{2g}0;1m\rangle$,	where$\mu =\,0,\,2$, $\nu =\,\pm\,\frac{1}{2}$, $m\,=\,0,\,\pm 1$ ($|E_g0\rangle \propto d_{z^2}; |E_g2\rangle \propto d_{x^2-y^2}$), and address the low-energy physics for nickelates formed by a system of such octets. This approach will allow taking into account the effects of competition of various degrees of freedom in the most general way.

\section{Effective model Hamiltonian}

\begin{figure}
	\includegraphics[width=\linewidth]{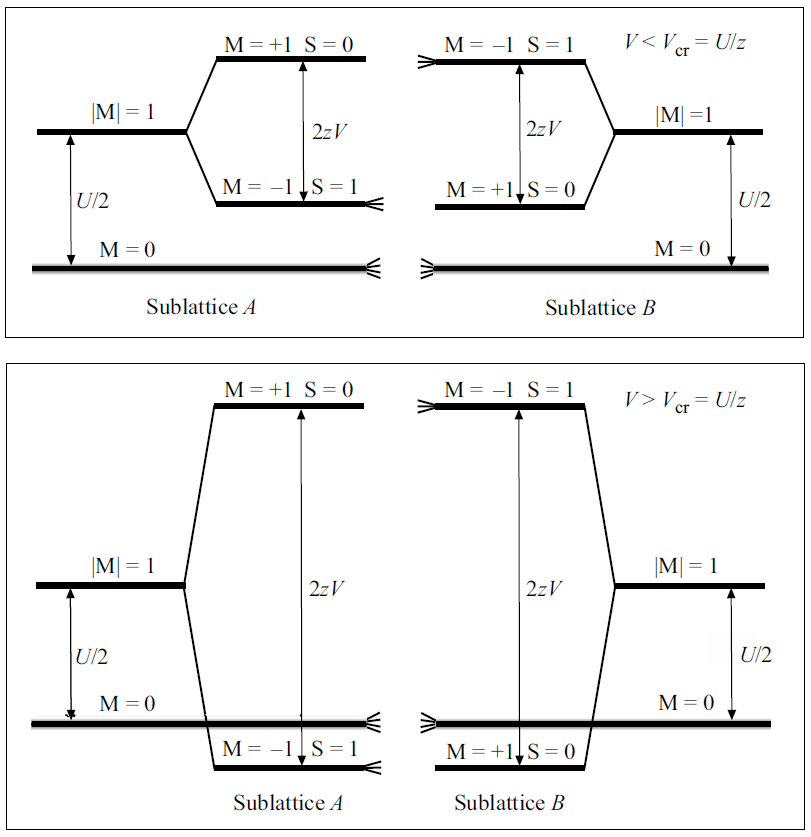}
	\caption{Energy spectrum scheme of the NiO6 cluster octets in two model orthonickelate sublattices in the ground state: upper panel  $V<V_{cr}$, , lower panel $V>V_{cr}$.
	}
	\label{fig1}
\end{figure}

Describing orthonickelate as a pseudospin-orbital-spin octet system is an extremely difficult problem. We have to use additional simplifications in real practice on the basis of the existing experimental data. Within a minimal model, our focus will be made on the only one relevant charge degree of freedom neglecting electron-lattice and superexchange interactions, and also vibronically reduced single-particle transport effects. In this approximation, the local octet structure will be reduced to the fourfold degenerate JT level with $M = 0$, singlet with $M = +1$ and spin triplet with $M = ?1$.	Assumption on the ''inactivity'' of the Jahn-Teller quartet is based on the absence, both above and below the MIT temperature, of any significant traces of Jahn-Teller distortions of the NiO$_6$ octahedra and orbital ordering [25], which implicitly indicates the ''anti Jahn-Teller'' disproportionation effect in the ground state of nickelates [26]. This allows splitting of the Jahn-Teller spin-orbital quartets with $M = 0$ to be neglected, i. e. the ''Jahn-Teller physics'' to be actually neglected. Taking into account all approximations, the effective Hamiltonian describing the relevant charge degree of freedom may be represented as a sum of three main contributions
\begin{equation}
	{\hat H} =	{\hat H}_{loc} +{\hat H}_{nloc}+{\hat H}_{tr}^{(2)} \, ,
	\label{loc+}	
\end{equation}
including the potential energy of local and non-local correlations and the two-particle transfer energy --- kinetic energy of effective composite bosons. This minimal model may be considered as some ''zero'' approximation that takes into account the leading charge degree of freedom.

Effective Hamiltonian of the system of noninteracting NiO$_6$ centers includes only local correlations
\begin{equation}
	{\hat H}_{loc} = \frac{U}{2}\sum_{i} {\hat \Sigma}_{iz}^2
	\label{loc}	
\end{equation}
--- equivalent to the single-ion axial spin anisotropy describing the bare pseudospin splitting effects. Positive values of the local correlation parameter $U > 0$ stabilize the spin-orbital JT quartet $|10;E_g\mu;\frac{1}{2}\nu\rangle$ that consists of the [NiO$_6$]$^{9-}$-centers corresponding to the pseudospin projection $M = 0$, while the negative values of $U < 0$ stabilize the disproportionation system of spin-charge [NiO$_6$]$^{10-,8-}$-centers corresponding to the pseudospin projection $M = \pm1$. In accordance with the experimental data regarding the observed JT effect for well isolated Ni$^{3+}$ ions in LaAlO$_3$; Ni$^{3+}$ [27,28], positive sign for $U$ in nickelates is selected below. However, with positive $U$, disproportionation is possible only with a quite high screened Coulomb inter-center interaction, or non-local correlations, described by the effective Hamiltonian
\begin{equation}
	{\hat H}_{nloc} =  \frac{1}{2}V\sum_{i\neq j} {\hat \Sigma}_{iz}{\hat \Sigma}_{jz}\, ,
	\label{nloc}	
\end{equation}
--- equivalent of two-ion spin anisotropy or Ising exchange. Non-local correlations drive the classical disproportionation (''site-centered'' charge order or CO phase) with G-type ordering of spin-triplet electron and spinless hole centers, which corresponds to the paramagnetic phase when only nn-interactions of the nearest neighbors are accounted for. 

Effective two-particle transfer Hamiltonian
\begin{equation}
	{\hat H}_{tr}^{(2)}=-\frac{1}{2}t_b\sum_{i\not= j} \left({\hat \Sigma}_{i+}^{2}{\hat \Sigma}_{j-}^{2}+ {\hat \Sigma}_{i-}^{2}{\hat \Sigma}_{j+}^{2}\right)
	\label{Hkin2}
\end{equation}is equivalent to the transfer Hamiltonian of effective composite two-electron spin-triplet bosons with the $e_g^2$;${^3}A_{2g}$ configuration; and the transfer integral $t_b$.

By introducing the creation/annihilation operators ${\hat B}_{\mu}^{\dagger}$/${\hat B}_{\mu}$ for the effective composite boson and selecting the spin component $\mu=0,\pm 1$, we rewrite ${\hat H}_{tr}^{(2)}$ as follows
\begin{equation}
	{\hat H}_{tr}^{(2)}=-t_b\sum_{i\not= j,\mu} {\hat B}_{i\mu}^{\dagger}{\hat B}_{j\mu} \, .
	\label{Hb}
\end{equation}
Unlike the classical correlation contributions (3) and (4), the quantum transfer operator ${\hat H}_{tr}^{(2)}$ doesn't preserve the local pseudospin projection $\Sigma_{iz}$, i. e. the local charge state. In the molecular filed approximation, this operator leads to formation of local quantum superpositions [17,18]
\begin{equation}
	|\alpha\rangle = \cos\alpha\,|+1\rangle + \sin\alpha\,|-1\rangle \, ,
	\label{+-}
\end{equation}
where $\langle\Sigma_z\rangle=\cos2\alpha=\delta$. It is natural that quantum superpositions (7) with $|\delta |<\,1$ differ fundamentally from the classical states with the corresponding charge density. Thus, when $\delta=0$, we deal with the local superposition of the Ni$^{2+}$- and Ni$^{4+}$-centers, rather than with the Ni$^{3+}$-center. To distinguish classical and quantum states with formally identical value of $\delta$, we can use a value of the local order parameter $\langle\Sigma^2_z\rangle$ equal to 1 for any superposition (7) and equal to zero for the Ni$^{3+}$-center corresponding to $M = 0$.

Consideration of the two-particle transport leads to charge density transfer with mixing the local charge states with $M=\pm1$, occurrence of uncertainty of the charge state of the NiO$_6$ clusters with mean charge (valence) [NiO$_6$]$^{9\pm\delta}$ (Ni$^{3\pm\delta}$), and formation of the quantum disproportionation CDq-phase.

Transfer of the effective composite spin-triplet boson corresponds to the transfer of both charge and spin density with preserving the conventional spin, but with appearance of uncertainty of the local spin value so that ${\hat H}_{tr}^{(2)}$ is actually also a nontraditional spin operator or a double boson exchange similar to the traditional Zener double exchange [29-31]. However, this spin dependence is nontrivial. The Hamiltonian initiating transport is spinless, so ${\hat H}_{tr}^{(2)}$ may be represented in a semiclassical approximation [30,32] as
\begin{equation}
	{\hat H}_{tr}^{(2)}=-t_b\sum_{i\not=j}  S_{ij}{\hat B}_{i}^{\dagger}{\hat B}_{j} \, ,
	\label{Hb1}
\end{equation}
where $S_{ij}$ is the spin function overlap integral in the common coordinate system that may be expressed in the simplest case through the angle $\theta_{ij}$ between the spin/magnetic moments ${\bf S}_i$ and ${\bf S}_j$ [17]:
\begin{equation}
	S_{ij}=\cos^2\frac{\theta_{ij}}{2}\, .
\end{equation}
$S_{ij}$ is obviously maximum for the ferromagnetic orientation of magnetic moments of neighboring sites, which is traditionally associated with the ferromagnetic nature of double exchange and attempts to introduce the Heisenberg-type effective spin Hamiltonian. However, the transfer Hamiltonian doesn't allow charge and spin degrees of freedom to be split. Appearance of the quantum uncertainty of a local spin value with local spin density in superpositions (7)
\begin{equation}
	\rho_s=\sin^2\alpha=\frac{1\pm |\delta|}{2} 
	\label{rhos}
\end{equation}
indicates that it is fundamentally impossible to associate the transfer operator with the effective spin Hamiltonian as is often the case in the traditional (''one-particle'') Zener double exchange theory [29-32].
 
Thus, unlike nonlocal correlations, the two-particle, or boson, transfer drives the formation of the quantum ferromagnetic CDq-phase with mean, but quantum-mechanical uncertain values of charge and spin for the NiO$_6$-centers described by quantum superpositions (7).
        
The prospect of forming unique phase states such as spin-triplet superconductivity or ''supersolid'' [33] is probably the most amazing detail of the quantum transport of composite spin-triplet bosons in the CDq- phase. In this context, note the recently discovered superconducting properties in mixed-valence La$_3$Ni$_2$O$_7$ [34].

\section{Effective field theory}

\begin{figure*}
	\centering
	\includegraphics[width=\linewidth]{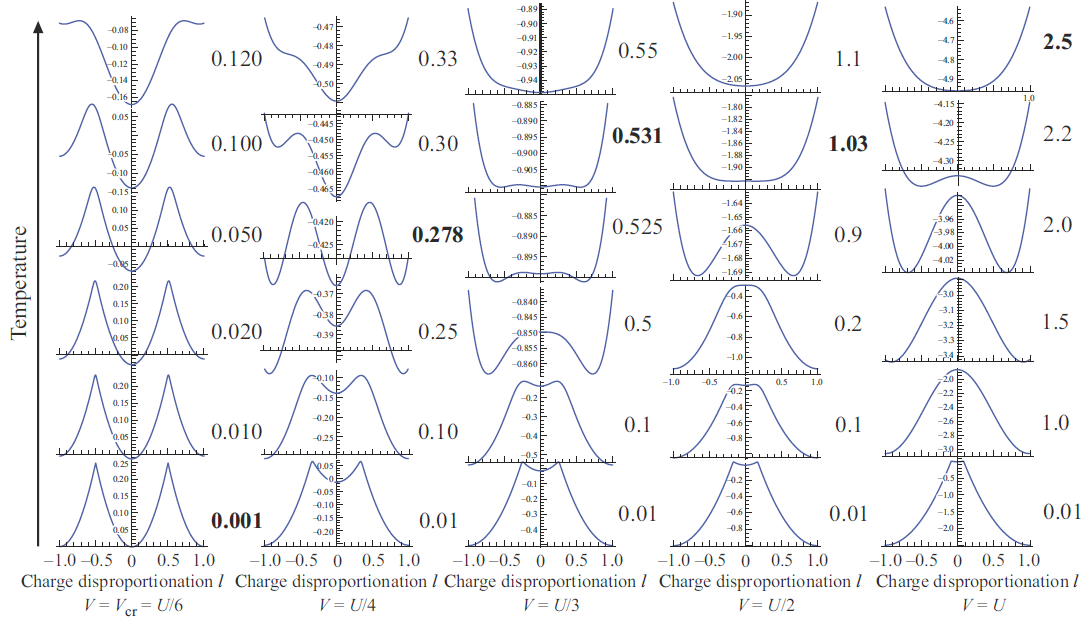}
	\caption{ Dependence of the free energy of the model nickelate (per site) on the order parameter $l$ at various temperatures and relations between the local and nonlocal correlation parameters.  }
	\label{fig2} 
\end{figure*}

The charge triplet model and pseudospin formalism indicate that it is possible to describe charge states in nickelates as well as in other JT magnets using methods that are well known in the spin magnet theory, primarily the simple effective field theory (EF) -- generalization of the mean-field theory -- which is a good starting point for physically clear semiquantitative description of strongly correlated systems.

The effective field approximation precisely accounts for all local (single-center) interactions, and all inter-center interactions are accounted for within the mean (molecular) field (MFA) theory typical for describing ''classical'' strongly correlated spin-magnetic systems.

Assuming the existence of two interpenetrating sublattices ($A$ and $B$) in the CO phase of orthonickelate, we introduce two charge order parameters of the ferro- and antiferro-type, respectively:
$$
\Delta n = \frac{1}{2}(\langle {\hat \Sigma}_{zA}\rangle + \langle {\hat \Sigma}_{zB}\rangle )
$$
($\Delta n$  is the deviation from half-occupancy) and
$$ 
l=\frac{1}{2}(\langle {\hat \Sigma}_{zA}\rangle - \langle {\hat \Sigma}_{zB}\rangle ) \, ,
$$
where $\langle {\hat \Sigma}_{zA,B}\rangle $ are the local order parameters $\langle {\hat \Sigma}_{z}\rangle$ for sublattices $A, B$.

In the molecular field approximation for the bilinear Hamiltonian of nonlocal charge correlations:
$$
\frac{1}{2} \sum_{i\neq j}V_{ij} {\hat \Sigma}_{zi} {\hat \Sigma}_{zj} \simeq   \frac{1}{2} \sum_{i\neq j}2V_{ij} {\hat \Sigma}_{zi}  
\langle {\hat \Sigma}_{zj}\rangle -\frac{1}{2} \sum_{i\neq j}V_{ij} \langle {\hat \Sigma}_{zi}\rangle  \langle {\hat \Sigma}_{zj}\rangle
$$	 
\begin{equation}
	=-\sum_{i}h_i{\hat \Sigma}_{zi} 	-\frac{1}{2} \sum_{i\neq j}V_{ij} \langle {\hat \Sigma}_{zi}\rangle  \langle {\hat \Sigma}_{zj}\rangle  \, ,
	\label{MFA}
\end{equation}
where
\begin{equation}
	h_i= - \sum_{j\neq i}V_{ij} \langle {\hat \Sigma}_{zj}\rangle  \, 
\end{equation}
is the molecular field. The last ''non-operator'' term in (11) that is fully dependent on pseudospin mean values shall be nevertheless included both in the octet spectrum and free energy.

The effective Hamiltonian may be represented as a sum of single-center contributions
\begin{equation}
	\mathcal{H}_0 = \sum_{c=1}^{N/2} \mathcal{H}_c 
	,\qquad
	\mathcal{H}_c = \mathcal{H}_A + \mathcal{H}_B
	,
\end{equation}
where 
\begin{equation}
	\mathcal{H}_\alpha = \frac{U}{2} {\hat \Sigma}_{z\alpha}^2  -  h_{\alpha} {\hat \Sigma}_{z\alpha} \, ,
	\label{HAB}
\end{equation}
$\alpha=A,B$, 
\begin{equation}
	h_{A,B}=-zV\langle {\hat \Sigma}_{zB,A}\rangle
	\label{h}
\end{equation}
are molecular fields in the nearest neighbor approximation, $z = 6$ is the number of the nearest neighbors.

Interestingly, $\mathcal{H}_\alpha$ resembles the Hamiltonian of the spin $S = 1$ center with the ''easy-plane'' type axial single-ion anisotropy in the external field oriented along the symmetry axis.

Figure 1 shows the model energy spectrum of the Nicenter octets in two model nickelate sublattices in atomic limit with $U > 0$. It is obvious that with $V<V_{cr}=U/z$, where $z$ is the coordination number ($z= 6$ for cubic perovskite), the ground state of Ni-centers corresponds to the JT quartet (see the upper panel), and with $V > V_{cr}$ the ground state of Ni-centers corresponds to the classical disproportionated CO configuration with electron (hole) centers in sublattice $A(B)$ (see the lower panel). Spectrum in Figure 1 is used to find the single-particle and two-particle optical transition energies with charge-transfer in the CO phase of nickelates ($\Delta_1=zV-U$ and $\Delta_2=4zV$, respectively) and to evaluate the local and nonlocal correlation parameters. Unfortunately, there is extremely scarce data for nickelates [35].

Octet distribution function is written as
$$
Z_c = \mathrm{Tr\,} \left( e^{-\beta \mathcal{H}_c } \right)
= \mathrm{Tr\,} \left( e^{-\beta \mathcal{H}_A } \right) \; \mathrm{Tr\,} \left( e^{-\beta \mathcal{H}_B } \right)
= Z_A Z_B \, 	,
$$
where $\beta=1/k_BT$, 
\begin{equation}
	Z_{A,B}=4+e^{-\frac{1}{2}\beta U}(3e^{-\beta h_{A,B}}+e^{\beta h_{A,B}}) .
\end{equation}

Free energy per center is written as
\begin{equation}
	f = -\frac{1}{2\beta} (\ln Z_A+\ln Z_B) -\frac{1}{2}zV(\Delta n^2-l^2) \, .
	\label{f}
\end{equation}
By minimizing the free energy, equations may be derived to determine the order parameters.

\begin{figure}
	\centering
	\includegraphics[width=\linewidth]{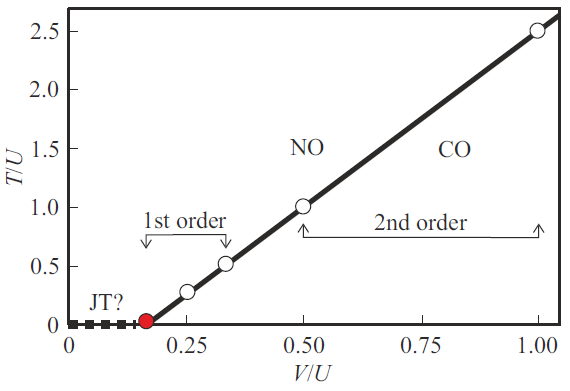}
	\caption{ Dependence of the CO-NO transition temperature on non-local correlation parameter in the atomic limit of the model ($T-V$-phase diagram). }
	\label{fig3}
\end{figure}

\section{CO-NO transition in the model nickelate: atomic limit}

For illustration, consider the simplest atomic limit of the minimal nickelate model where the CO-NO insulator-metal transition and phase diagram will be defined by only one parameter --- the $V/U$ ratio of non-local and local correlations. Figure 2 shows the dependence of free energy $f(l)$ on $l$ for various temperatures and values of $V$ in terms of $U > 0$. Figure 3 shows the phase $T$-$V$ diagram of the model nickelate in the atomic limit ($U$-$V$-model). Here, line $T(V)$ separates the low-temperature insulating CO-phase with the classical charge disproportionation and the high temperature disordered NO-phase that is associated with the bad-metallic phase.

CO-NO transition takes place only when $V$>$V_{cr}$\,=\,$\frac{1}{z}U$\,=\,$\frac{1}{6}U$, so, when $V$\,=\,$V_{cr}$ and $T = 0$, free energies of the CO and NO phases coincide, the transition temperature turns into zero, $T_{CO} = 0$. When $V < Vcr$, cooperative JT ordering may be implemented. When $V > V_{cr}$, the CO-NO temperature grows almost linearly as $V$ increases. Analysis of the $f (l)$ dependence shows that with $V_{cr}$<$V$$\leq$\,$\frac{1}{3}U$ it is typical of first-order phase transitions and with $V$$\geq$\,$\frac{1}{2}U$ it is typical of second-order phase transitions. In the intermediate region $\frac{1}{3}U$<$V$<\,$\frac{1}{2}U$, features typical both of the second-order and first-order phase transitions are observed. High temperature NO-phase is stable throughout the temperature range up to the lowest temperatures. The low-temperature insulating CO-phase simultaneously becomes stable above the critical temperature $T_{CO}$, i. e. the maximum temperature at which the free energies of the NO and CO phases coincide.

Features of coexistence of the disordered NO-phase and ordered CO-phase indicate phase separation as a typical state of the model orthonickelate, which automatically explains the unique sensitivity of orthonickelates to nonstoichiometry, sample shape (bulk/film), real and chemical pressure, and isotopic substitution [6].

Surprisingly, but the addressed simple purely electronic $U$-$V$ model makes it possible to describe a set of fundamental features of the spontaneous insulator-bad metal transition associated with the CO-NO transition of insulating CO-phase ''melting'' in orthonickelates. Thus, the first-order nature of transition is easily explained on the assumption of relatively low values of the non-local correlation parameter $\frac{1}{6}U\leq V\leq\frac{1}{3}U$. Universal nature of phase separation in nickelates [36-39] typical of strongly correlated systems [40] is naturally associated with a wide temperature region of coexistence of the bad-metallic NO-phase and insulating CO-phase. Transition temperature behavior in the rare-earth orthonickelate series can be easily associated with the nonlocal correlation parameter variation. Indeed, an increase in the Ni-O-Ni coupling angle and a corresponding increase in the Ni-Ni distance in the LuNiO$_3$--LaNiO$_3$ series lead to a decrease in the V parameter of non-local correlations and, in accordance with the predictions of the purely electronic model, to a decrease in the insulator-metal transition temperature, which is observed experimentally (see, for example, [1-3,41]). It is not unlikely that in LaNiO$_3$ (LNO) the non-local correlation parameter approaches the critical value, so the ground state in this nickelate may result from the phase separation -- a bad-metallic disordered NO-phase co-existing with the CO-phase. The neutron diffraction data actually show that on the approach to 0 K the crystal symmetry of LNO becomes two-sublattice and locally monoclinic [42]. This indicates that, in spite of metallicity, the system is very close to the insulating state with charge disproportionation and monoclinic global structure. This generally explains a set of anomalous properties of LNO, in particular, both strong paramagnetism and antiferromagnetic correlations [42-45]. Authors of [46] assume that LNO is a quantum-critical metal close to the antiferromagnetic quantum critical point (QCP). Moreover, according to [44], a combination of metallic properties and antiferromagnetic ordering which is quite rare for transition metal oxides, was found in LaNiO$_3$ single crystals with a relatively high Neel temperature: $T_N = 157$ K and ${\bf Q}_{AFM}=(1/2,0,1/2)$, typical for the entire series of orthonickelates.

The minimal model of the spontaneous NO-CO transition, that is, the model of the formation of long-range CO ordering with decreasing temperature, assumes the formation of short-range order, in fact of electron-hole (EH) centers, at temperatures significantly exceeding TCO. These EH centers, that are relatively stable due to strong electron-lattice interaction, form a kind of bipolarons that are	''nucleation centers'' of the low-temperature disproportionation CO-phase. Thus, evolution of the NO-phase in our model includes several stages from a totally disordered ''octet'' system with mixed valence of the Debye-Huckel type NiO$_6$ centers and strongly screened Coulomb interactions to non-degenerate ''bipolaron gas'', ''bipolaron liquid'' and, finally, condensation of the phase into an ordered disproportionated CO-phase with $T$\,=\,$T_{CO}$. It is this scenario proposed by N.F. Mott many years ago for a high-temperature badmetallic phase in VO$_2$ [5] that is validated by numerous experiments for orthonickelates, primarily by the data [14,16]. Summing up, an emphasis shall be made again on the similarity of the model Hamiltonian of local and nonlocal correlations with the Hamiltonian of an Ising $S = 1$ antiferromagnet with strong easy-plane uniaxial single-ion anisotropy in the external field oriented along the axis of symmetry.

\section{Two-particle transfer and formation of the quantum disproportionation CDq phase}

\begin{figure*}
	\centering
	\includegraphics[width=0.65\linewidth]{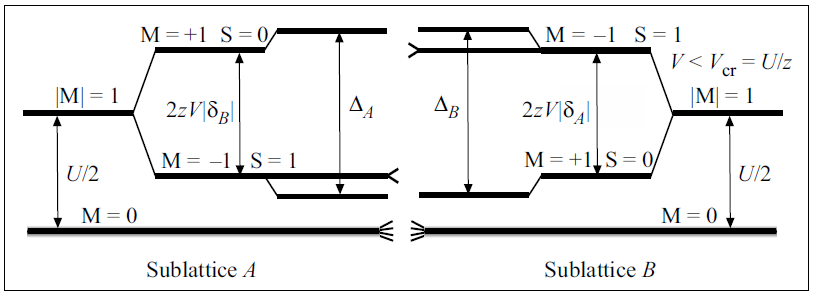}
	\caption{Energy spectrum scheme of the NiO$_6$ cluster octets in two model orthonickelate sublattices in the ground state taking into account the two-particle transfer effect with $V<V_{cr}$. }
	\label{fig4}
\end{figure*} 

\begin{figure*}
	\centering
	\includegraphics[width=0.75\linewidth]{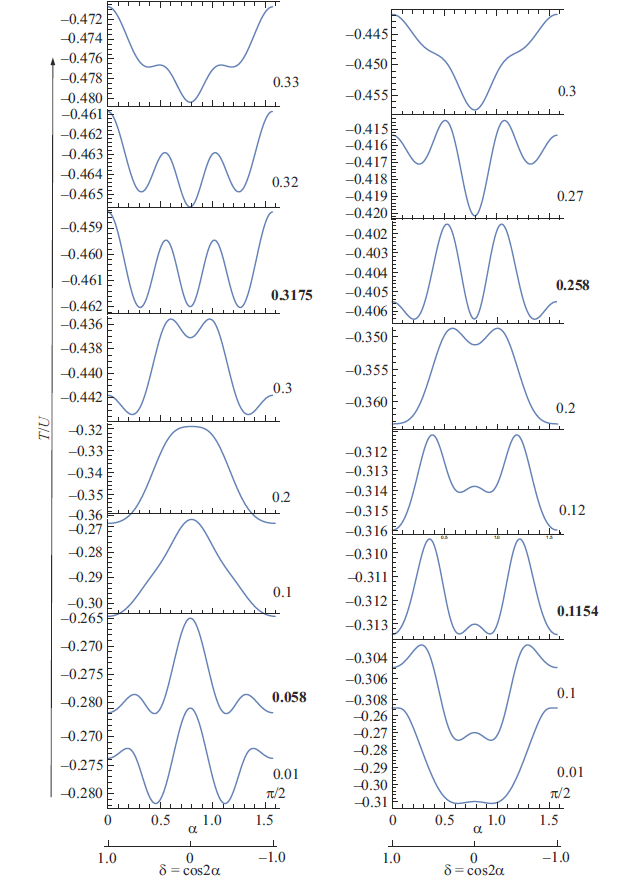}
	\caption{Dependence of the free energy of the model nickelate on $l$ with various temperatures with  $V$\,=\,0.25\,$U$, $t_b$\,=\,0.17\,$U$ (left) and $t_b$\,=\,0.18\,$U$ (right). CDq-CO and CO-NO transition temperatures, respectively, are shown in bold.}
	\label{fig5}
\end{figure*}

Complete certainty of the local charge state of the NiO$_6$ clusters, or of the pseudospin projection $\Sigma_z$, is the main distinguishing feature of the atomic limit. The ground state of the two-sublattice model with $V$\,>\,$V_{cr}$ resembles the Neel state of a G-type spin antiferromagnet and is formed by the system of [NiO$_6$]$^{10-}$ ($\langle\Sigma_z\rangle=-1$) centers in one sublattice and [NiO$_6$]$^{8-}$ ($\langle\Sigma_z\rangle=+1$) centers in another sublattice with $S = 1$ and $S = 0$, respectively. However, these conclusions disagree with magnetic neutron diffraction data [17-19] indicating redistribution of spin and, thus, of charge density between sublattices with formation of local charge states with [NiO$_6$]$^{(9-\delta )-}$ type mixed valence, or simply Ni$^{3-\delta}$ ($-1\leq\delta\leq +1$) representing local quantum superpositions (7).

Unlike the atomic limit with actually classical nonlocal inter-center correlation effect, the introduction of a molecular field for describing the quantum effect of charge transfer (quantum tunneling) seems to be a rougher approximation, but, as we hope, not lacking in the potential of qualitative and semi-quantitative predictions. Within MFA, the transfer Hamiltonian ${\hat H}_{tr}^{(2)}$ with a fixed conventional spin component is
\begin{multline}
{\hat H}_{tr}^{(2)}=-\sum_{i\not= j} t_{ij}^b \left({\hat \Sigma}_{i+}^{2}\langle{\hat \Sigma}_{j-}^{2}\rangle + {\hat \Sigma}_{i-}^{2}\langle{\hat \Sigma}_{j+}^{2}\rangle\right) +
\\
	\frac{1}{2}\sum_{i\not= j} t_{ij}^b \left(\langle{\hat \Sigma}_{i+}^{2}\rangle\langle{\hat \Sigma}_{j-}^{2}\rangle + \langle{\hat \Sigma}_{i-}^{2}\rangle\langle{\hat \Sigma}_{j+}^{2}\rangle\right)\,, 
	\label{Hkin2MFA}
\end{multline} 
where for quantum-mechanical mean values
\begin{equation}
	\langle{\hat \Sigma}_{i\pm}^{2}\rangle = \langle \alpha |{\hat \Sigma}_{i\pm}^{2}|\alpha\rangle = \frac{1}{2}\sin2\alpha = \frac{1}{2}\sqrt{1-\delta^2}\, .
\end{equation}
According to the effective field theory for the two-sublattice model, the operator part of Hamiltonian (18) is included in H? (14), which leads to modification of the octet energy spectrum consisting of two sublattices as shown in Figure 4, where
\begin{multline}
\Delta_{A,B}=2z\sqrt{V^2\delta_{B,A}^2+t_b^2(1-\delta_{B,A}^2)} \, =
\\
	=2z\sqrt{V^2\cos^2(2\alpha_{B,A})+t_b^2\sin^2(2\alpha_{B,A})} \, ,
\end{multline}
and tb is the transfer integral for the nearest neighbors.

For octet distribution functions $Z_{A,B}$ in sublattices, we obtain
\begin{equation}
	Z_{A,B}=4+e^{-\frac{1}{2}\beta U} \left( 2e^{-\beta h_{A,B}}+2\cosh{\frac{\Delta_{A,B}}{2}} \right) 
\end{equation}
Free energy per center is written as
\begin{equation}
	f = -\frac{1}{2\beta} \left( \ln Z_A+\ln Z_B \right)  +\frac{1}{2}zVl^2 + \frac{1}{4}zt(1-l^2) \, .
	\label{f2}
\end{equation}

\begin{figure*}
	\centering
	\includegraphics[width=\linewidth]{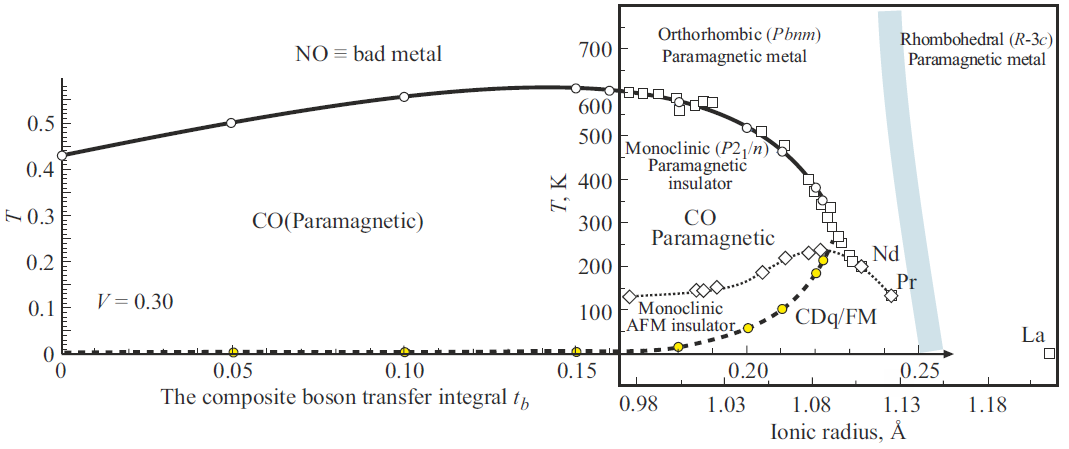}
	\caption{Model $T$-$t_b$-phase diagram for nickelates. Dependence of critical CDq-CO transition temperatures (filled circles and bold dashed line) and CO-NO (empty circles and bold solid line) on the magnitude of composite boson transfer integral $t_b$ and comparison with the experimental data for $T_{\mbox{{\footnotesize MIT}}}$ (squares) and $T_{\mbox{{\footnotesize N}}}$ (rhombs and dashed line) with indication of crystalline, electronic and magnetic structures for various orthonickelates RNiO$_3$ [3]. The curves were plotted by eye, $T$, $V$ and $t_b$ are given in terms of $U$.}
	\label{fig6}
\end{figure*}

Inclusion of the quantum effect of composite boson transfer ${\hat H}_{tr}^{(2)}$ ($U$-$V$-$t_b$-model) together with the local and non-local correlations leads to fundamental restructuring of the ground state, phase transitions and temperature phase diagram of nickelates. With quite low values of tb compared with $V$, the quantum CDq phase existence temperature range turns to be small, so a small increase in temperature leads to the CDq-CO phase transition to the classical disproportionation phase with further CO-NO transition to the bad-metallic NO phase. However, competition of potential and kinetic energies of effective composite bosons, i. e non-local correlations and two-particle transfer, leads to an unexpected effect. With growth of the boson transfer integral, the CDq-CO transition temperature $T_{\mbox{{\footnotesize CDq}}}$ first grows slowly being close to zero, and then starts growing rapidly at some critical value of $t_b^s$ , while $T_{\mbox{{\footnotesize CO}}}$ of the CO-NO transition to the bad-metallic phase, i. e. the insulator-metal transition temperature, first grows slowly, achieves its maximum and then starts decreasing rapidly at $t_b\approx t_b^s$, and $T_{\mbox{{\footnotesize CO}}}$ and $T_{\mbox{{\footnotesize CDq}}}$ become equal at a critical value of $t_b^*$, and only the CDq phase survives at $t_b$\,>\,$t_b^*$.  |$\delta$| that defines local quantum superpositions (7) grows slowly as $t_b$ increases in a wide range while remaining close to 1, and then starts decreasing rapidly at $t_b\sim t_b^s$ and vanishes at $t_b\sim t_b^*$.

All these features are well illustrated in Figure 6 that shows phase diagram $T$-$t_b$ at $V$\,=\,0.30\,$U$ corresponding to a poorly pronounced first-order CO-NO transition region. With the simplest assumption on the linear dependence of $t_b$ on the ionic radius of R-ions in the range of its real values for orthonickelates, $T(t_b)$ perfectly reproduces the R-dependence of paramagnetic insulator--bad metal transition temperatures for RNiO$_3$ with $T_{\mbox{{\footnotesize MIT}}}$\,$\neq$\,$T_{\mbox{{\footnotesize N}}}$ (R = Lu, $\ldots$, Sm) [3] at $U$$\approx$\,1000 K, which immediately leads to $V$$\approx$\,300 K. In this case, the boson transfer integral varies from $t_b$\,$\approx$\,165 K corresponding to LuNiO$_3$ to $t_b$\,=\,$t_b^*$\,$\approx$\,225 K approximately corresponding to SmNiO$_3$, which is in good qualitative and quantitative agreement with theoretical predictions [18].

With $t_b$\,>\,$t_b^*$, our model, in line with the experiment, indicates fundamental variation of the MIT behavior from NO $\rightarrow$ classical paramagnetic CO-phase to NO $\rightarrow$ quantum magnetic CDq-phase. However, the observed complex noncollinear antiferromagnetic structure ${\bf Q}_{AFM}$\,=\,(1/2,0,1/2) [1,2] in this phase and unusual dependence $T_{\mbox{{\footnotesize MIT}}}$(R)\,=\,$T_{\mbox{{\footnotesize N}}}$(R) for R = Nd, Pr and similar compounds indicate the limitation of the minimal purely charge model and the need to consider rather unusual competition of the ferromagnetic boson double exchange and antiferromagnetic superexchange, i. e. transition to the $U$-$V$-$tb$-$J$-model. A more realistic model shall also consider a considerable decrease in the $V$ in the LuNiO$_3$--LaNiO$_3$ series.

\section{Conclusion}

A new purely electronic scenario is presented for the insulator--bad metal transition in RNiO$_3$ based on the minimal $U$-$V$-$t_b$-model that takes into account the only one charge degree of freedom within the charge triplet model, pseudospin formalism and effective field theory, which ensures a physically clear and mathematically trivial description typical of traditional spin-magnetic systems. Despite the simplicity of the model, it reproduces the main features of the $T$-R-phase diagrams for RNiO$_3$ with physically validated estimates of the $U$-$V$-$t_b$-model parameters. High-temperature bad-metallic phase is associated with the disordered mixed-valence NO-phase of the charge triplet system, and the insulating disproportionated CDphase includes the low-temperature quantum ferromagnetic CDq-phase and high-temperature classical paramagnetic (in the nearest neighbor model!) CO-phase separated by the CDq-CO phase transition (R = Lu, $\ldots$, Sm), or only the magnetic CDq-phase (R = Nd, Pr). For detailed description of the quantum CDq-phase, it is necessary to go beyond the purely charge model taking into account the unusual competition of the ferromagnetic bosonic double exchange and antiferromagnetic superexchange Ni$^{2+}$-O$^{2-}$-Ni$^{2+}$. The developed pseudospin formalism makes it possible to effectively consider the electron-lattice interaction, primarily the essential contribution of a so-called 	''breathing'' mode of local distortions of the NiO$_6$ centers and bipolaron effects as well as other signs of the important role of electron-lattice coupling, which are experimentally observed in nickelates [2].

\section*{Funding}
The study was supported by the Ministry of Science and Higher Education of the Russian Federation under project FEUZ-2023-0017.

\section*{References}
{\setlength{\parindent}{0cm}
	\parskip=0.5em

[1]	M.L. Medarde. J. Phys.: Condens. Matter \textbf{9}, 1679 (1997).

[2]	S. Catalano, M. Gibert, J. Fowlie, J. Iniguez, J.-M. Triscone, J. Kreisel. Rep. Prog. Phys. \textbf{81}, 046501 (2018).

[3]	D.J. Gawryluk, Y.M. Klein, T. Shang, D. Sheptyakov, L. Keller, N. Casati, Ph. Lacorre, M.T. Fernandez-Diaz, J. RodriguezCarvajal, M. Medarde. Phys. Rev. B \textbf{100}, 205137 (2019).

[4]	M. Imada, A. Fujimori, Y. Tokura. Rev. Mod. Phys. \textbf{70}, 1039 (1998).

[5]	N.F. Mott. Rev. Mod. Phys. \textbf{40}, 677 (1968).

[6]	M. Medarde, P. Lacorre, K. Conder, F. Fauth, A. Furrer. Phys. Rev. Lett. \textbf{80}, 2397 (1998).

[7]	J.-S. Zhou, J.B. Goodenough, B. Dabrowski. Phys. Rev. Lett. \textbf{94}, 226602 (2005).

[8]	M. Tyunina, M. Savinov, O. Pacherova, A. Dejneka. Sci Rep \textbf{13}, 12493 (2023).

[9]	R. Jaramillo, S. Ha, D. Silevitch, S. Ramanathan. Nature Phys \textbf{10}, 304 (2014).

[10] S.D. Ha, R. Jaramillo, D.M. Silevitch, F. Schoofs, K. Kerman, J.D. Baniecki, S. Ramanathan. Phys. Rev. B \textbf{87}, 125150 (2013).

[11]	A.J. Hauser, E. Mikheev, N.E. Moreno, T.A. Cain, J. Hwang, J.Y. Zhang, S. Stemmer. Appl. Phys. Lett. \textbf{103}, 182105 (2013).

[12]	S.K. Ojha, S. Ray, T. Das, S. Middey, S. Sarkar, P. Mahadevan, Z. Wang, Y. Zhu, X. Liu, M. Kareev, J. Chakhalian. Phys. Rev. B \textbf{99}, 235153 (2019).

[13]	A. Stupakov, T. Kocourek, O. Pacherova, G. Suchaneck, A. Dejneka, M. Tyunina. Appl. Phys. Lett. \textbf{124}, 102103 (2024).

[14]	J.L. Garcia-Munoz, R. Mortimer, A. Llobet, J.A. Alonso, M.J. Martinez-Lope, S.P. Cottrell. Physica B: Condensed Matter \textbf{374}, 87 (2006).

[15]	C. Piamonteze, H.C.N. Tolentino, A.Y. Ramos, N.E. Massa, J.A. Alonso, M.J. Martinez-Lope, M.T. Casais. Phys. Rev. B \textbf{71}, 012104 (2005).

[16]	J. Shamblin, M. Heres, H. Zhou, J. Sangoro, M. Lang, J. Neuefeind, J.A. Alonso, S. Johnston. Nat Commun. \textbf{9}, 86 (2018).

[17]	A.S. Moskvin. JETP \textbf{167}, 3 (2025). pp. 115-131.

[18]	A.S. Moskvin. JETP Letters \textbf{121}, 6 (2025). pp. 411-420. 

[19] T.M. Rice, L. Sneddon. Phys. Rev. Lett. \textbf{47}, 689 (1981).

[20]	A.S. Moskvin. J. Phys.: Condens. Matter \textbf{25}, 085601 (2013).

[21]	A. Moskvin. Magnetochemistry \textbf{9}, 224 (2023).

[22]	A.S. Moskvin. Phys. Rev. B \textbf{84}, 075116 (2011).

[23]	A. Moskvin, Y. Panov. Condens. Matter \textbf{6}, 24 (2021).

[24]	A.S. Moskvin, Yu.D. Panov. JMMM \textbf{550}, 169004 (2022).

[25]	V. Scagnoli, U. Staub, A.M. Mulders, M. Janousch, G.I. Meijer, G. Hammerl, J.M. Tonnerre, N. Stojic. Phys. Rev. B \textbf{73}, 100409(R) (2006).

[26]	I.I. Mazin, D.I. Khomskii, R. Lengsdor, J.A. Alonso, W.G. Marshall, R.M. Ibberson, A. Podlesnyak, M.J. Martinez Lope, M.M. Abd-Elmeguid. Phys. Rev. Lett. \textbf{98}, 176406 (2007).

[27]	T. Ivanova, V. Petrashen, N. Chezhina, Y. Yablokov. Phys. Solid State \textbf{44}, 1468 (2002).

[28]	M.N. Sanz-Ortiz, F.O. Rodriguez, J. Rodriguez, G. Demazeau. J. Phys.: Condens. Matter \textbf{23}, 415501 (2011). 

[29] C. Zener. Phys. Rev. \textbf{82}, 403 (1951).

[30]	P.W. Anderson, H. Hasegawa. Phys. Rev. \textbf{100}, 675 (1955).

[31]	P.G. de Gennes. Phys. Rev. \textbf{118}, 141 (1960).

[32]	E. Muller-Hartmann, E. Dagotto. Phys. Rev. B \textbf{54}, R6819 (1996).

[33]	R. Micnas, J. Ranninger, S. Robaszkiewicz. Rev. Mod. Phys. \textbf{62}, 113 (1990).

[34]	H. Sun,	M. Huo,	X. Hu J. Li,	Z. Liu,	Y. Han,	L. Tang, Z. Mao, P. Yang, B. Wang, J. Cheng, D.-X. Yao, G.-M. Zhang, M. Wang. Nature \textbf{621}, 493 (2023).

[35]	J. Ruppen, J. Teyssier, O.E. Peil, S. Catalano, M. Gibert, J. Mravlje, J. Triscone, A. Georges, D. van der Marel. Phys. Rev. B \textbf{92}, 155145 (2015).

[36]	D. Preziosi, L. Lopez-Mir, X. Li, T. Cornelissen, J.H. Lee, F. Trier, K. Bouzehouane, S. Valencia, A. Gloter, A. Barthelemy, M. Bibes. Nano Lett. \textbf{18}, 2226 (2018).

[37]	K.W.	Post,	A.S.	McLeod,	M.	Hepting,	M.	Bluschke, Y. Wang, G. Cristiani, G. Logvenov, A. Charnukha, G.X. Ni, P. Radhakrishnan, M. Minola, A. Pasupathy, A.V. Boris, E. Benckiser, K.A. Dahmen, E.W. Carlson, B. Keimer, D.N. Basov. Nat. Phys. \textbf{14}, 1056 (2018).

[38]	J.H. Lee, F. Trier, T. Cornelissen, D. Preziosi, K. Bouzehouane, S. Fusil, S. Valencia, M. Bibes. Nano Letters \textbf{19}, 7801 (2019).

[39]	J. del Valle, R. Rocco, C. Dominguez, J. Fowlie, S. Gariglio, M.J. Rozenberg, J.-M. Triscone. Phys. Rev. B \textbf{104}, 165141 (2021).

[40]	D. Khomskii. Physica B: Condensed Matter \textbf{280}, 325 (2000),

[41]	Y.M. Klein, M. Kozlowski, A. Linden, P. Lacorre, M. Medarde. Crystal Growth and Design \textbf{21}, 4230 (2021).

[42]	B. Li, D. Louca, S. Yano, L.G. Marshall, J. Zhou, J.B. Goodenough. Adv. Electron. Mater. \textbf{2}, 1500261 (2015).

[43]	R. Scherwitz, S. Gariglio, M. Gabay, P. Zubko, M. Gibert, J.-M. Triscone. Phys. Rev. Lett. \textbf{106}, 246403 (2011).

[44]	H. Guo, Z.W. Li, L. Zhao, Z. Hu, C.F. Chang, C.Y. Kuo, W. Schmidt, A. Piovano, T.W. Pi, O. Sobolev, D.I. Khomskii, L.H. Tjeng, A.C. Komarek. Nat. Commun. \textbf{9}, 43 (2018). 

[45] A. Subedi. SciPost Phys. \textbf{5}, 020 (2018).

[46]	C.	Liu,	V.F.C.	Humbert,	T.M.	Bretz-Sullivan,	G.	Wang,
D. Hong, F. Wrobel, J. Zhang, J.D. Hoffman, J.E. Pearson, J.S. Jiang, C. Chang, A. Suslov, N. Mason, M.R. Norman, A. Bhattacharya. Nat. Commun. \textbf{11}, 1402 (2020).

[47]	J.L. Garcia-Munoz, J. Rodriguez-Carvajal, P. Lacorre. Phys. Rev. B \textbf{50}, 978 (1994).

[48]	J. Rodriguez-Carvajal, S. Rosenkranz, M. Medarde, P. Lacorre, M.T. Fernandez-Diaz, F. Fauth, V. Trounov. Phys. Rev. B \textbf{57}, 456 (1998).

[49]	M.T. Fernandez-Diaz, J.A. Alonso, M.J. Martinez-Lope, M.T. Casais, J.L. Garcia-Munoz. Phys. Rev. B \textbf{64}, 144417 (2001).

}

\end{document}